\documentclass[5p]{elsarticle}

\usepackage{amssymb}
\journal{TIPP09 Proceedings in NIMA}

\begin{document}

\begin{frontmatter}



\title{ATLAS monitored drift tube chambers for super\,-\,LHC}

%\author{}
%
%\address{}
%
%
%***************
%   Authors
%***************
%
\author[First]{Albert Engl}
\ead{albert.engl@physik.uni-muenchen.de}
\author[First]{Otmar Biebel}
%\cortext[cor1]{Corresponding author. Tel.: +0-000-000-0000; fax: +0-000-000-0000.} 
\author[First]{Ralf Hertenberger}
\author[First]{Alexander Mlynek}
\author[First]{Thomas A. Mueller}
\author[First]{Felix Rauscher}
%\author[]{}
%
%
%***************
%   Addresses
%***************
%
\address[First]     {LMU Munich, Am Coulombwall 1, 85748 Garching, Germany}
%\address[]{}
%\address[]{}

%
%===============================================================
%
%   Abstract
%
%===============================================================
%
%
\begin{abstract}

After the high-luminosity upgrade of the Large Hadron Collider (LHC) at CERN, 
the ATLAS muon spectrometer is expected to work at 10 times increased 
background rates of gammas and neutrons. This is challenging as the 
momentum resolution of the spectrometer is expected to be 10 \%.
This requires a single tube resolution of the muon drift tubes of 80 $\mu$m.
At background rates around 1000 Hz/cm$^{2}$ 
space charge effects will lead 
in the slow and non-linear AR:CO$_{2}$ = 93:7 gas mixture
to a degradation of the drift-tube
spatial resolution. This was studied before experimentally 
for gammas and low energetic
neutrons. Almost no information exists for fast neutrons.\\
As LHC did not yet reach the full proton energy of 7 TeV,
the real background rate at LHC standard luminosity is still unknown.
Therefore, we organized our studies under the following aspects:\\
- We investigated the influence of 11 MeV neutrons on the position resolution of 
  ATLAS MDT chambers.
  At flux densities between 4 and 16 kHz/cm$^{2}$, almost no influence on the 
  position resolution was found, it degrades by only 10 $\mu$m at  
  a detection efficiency of only 4$\cdot10^{-4}$.\\
- We investigated inert gas mixtures
  on fastness and linearity of their position-drifttime (r-t) relation.
  At a reduction of the maximum drift time by a factor of 2,
  the use of the present hardware and electronics might be possible.  
  For our experimental studies we used our Munich cosmic ray facility.
  It is equipped with 3 ATLAS BOS (Barrel Outer Small) MDT chambers and allows fast measurement
  of data sets with many million cosmic muon tracks. 
  Two gas mixtures show almost identical position
  resolution as the standard gas.
  The results are in good agreement with Garfield simulations.\\ 
- For spectrometer regions of highest background rates we contributed to the 
  investigation of newly developed 15 mm drift tubes. 
  Position resolutions have been measured as a function of gamma background rates
  between 0 and 1400 Hz/cm$^{2}$.\\
- Garfield simulations have been performed to simulate space charge effects
  due to gamma irradiation. Results will be presented for the standard geometry
  as well as for the new 15 mm drift tubes.
\end{abstract}

%
%
%===============================================================
%
%   Keywords
%
%===============================================================
%
%

\end{frontmatter}


%
%
%===============================================================
%
%   Main text

\section{Introduction}

	 Our gas- and detector-studies of ATLAS Monitored Drift Tube (MDT) muon detectors
	 are motivated by the requirements 
	 for the ATLAS muon spectrometer at future increased luminosities
	 of  the Large Hadron Collider (LHC) at CERN.
	 Luminosity upgrades are planned in several steps up to a final increase in luminosity
	 of about an order of magnitude at Super-LHC.
	 At the LHC design luminosity 10$^{34}$ cm$^{-2}$\,s$^{-1}$, background rates of gammas and neutrons 
	 of up to 10~cm$^{-2}$\,s$^{-1}$
	 are predicted for the barrel part of the spectrometer and rates up to
	 100~cm$^{-2}$\,s$^{-1}$ for the forward region of the endcap spectrometer.
	 
	 The ATLAS muon spectrometer will provide momentum resolution 
	 of 10 $\%$ at 1 TeV/c transverse muon momentum
	 at 3\,-\,6 Tm bending power generated by a superconducting air coil toroid.
	 It consists of 3 layers of precision drift tube chambers. An ATLAS MDT chamber\cite{ATLASTechreport} consists of
	 2 multilayers with 3 or 4 layers of 30 mm drift tubes each, which are filled with
	 an Argon and CO$_{2}$ gas mixture (93:7) at an absolute pressure of 3 bar. The voltage at the gold plated W-Re 
	 anode wire (diameter 50 $\mu m$) is 3080 V. These conditions correspond to a maximum
	 electron drift time of 700 ns and a gas gain of 2 $\cdot 10^4$.
	 
	 For sufficiently good sagitta reconstruction, 
	 single tube resolutions of around 80 $\mu m$ are envisaged. 
	 But at increased background rates of 1000 $\frac{Hz}{cm^2}$ or higher,
	 the single tube resolution deteriorates mostly due to 
	 space charge effects in the slow and nonlinear drift gas mixture. 
	 At too high occupancies unambigous track reconstruction
	 will become difficult or impossible\cite{Aleksadiss,Deile1,Deile2,Horvat}.
	 
	 This paper shows resolution measurements of fast and linear drift gases, the sensitivity of MDT chambers on fast neutrons, 
	 the influence of fast neutron background on the resolution,
	 and the comparison of space charge effects in 15 mm and 30 mm diameter drift tubes under $\gamma$\,-\,irradiation.

\section{Alternative drift gas mixtures}

	 The reduction of CO$_{2}$ from 7 to 3 \% reduces the maximum drift time by about 200 ns and the addition of small percentages of N$_{2}$ makes the
	 drift gas more linear.
	 Two inert gas mixtures (Ar-CO$_{2}$-N$_{2}$ 96:3:1\,/\,97:2:1) with linear position-drifttime relations (rt-relations) and maximum 
	 drifttimes around 450 ns are promising 
	 candidates. These mixtures were tested on their single tube resolutions.\\
	 At standard voltage and pressure conditions, the measurements were
	 performed at the cosmic ray facility in Garching (Fig.\,1).\\
	 A set of three MDT chambers is sandwiched by layers of
	 scintillation counters to trigger on cosmic muons. Using a 40 cm thick iron absorber, 
	 we selected muons with an energy higher than 600 MeV.\\
	 \begin{figure}[h]
	 \centerline{\includegraphics[width=7cm]{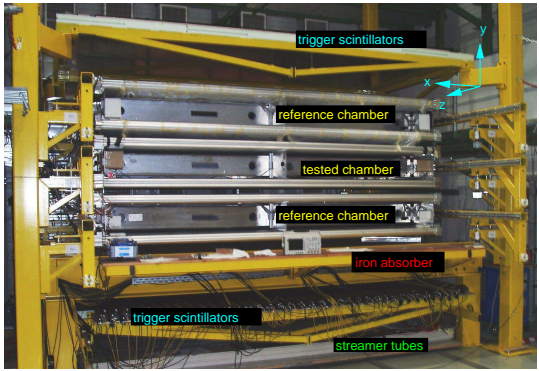}}
	 \caption{Cosmic ray facility in Garching (Munich)}
	 \end{figure}
	 The measured drifttime in each drifttube is transformed into a driftradius $r_{meas}$ with the help of the rt-relation.\\
	 The reference chambers provide a highly precise track reconstruction and therefore we get a precise prediction for the driftradius 
	 $r_{track}$ in the tubes of the test chamber. 
	 The difference between these two values defines the residual $res$:
	 \begin{equation}
	 res=r_{track}-r_{meas}.
	 \end{equation}
	 The width $\sigma$ of Gaussians describing the shape of the residual distribution of small drift radii intervals define the single tube resolution.
	 Fig.\,2 shows almost identical resolution for the tested mixtures including the standard gas. 
	 The most promising candidate is Ar-CO$_{2}$-N$_{2}$ in the ratio 96:3:1 \%. Quenching, afterpulsing and streamerrate of the alternative gas mixtures are ongoing.
	 \begin{figure}[h]
	 \centerline{\includegraphics[width=6cm]{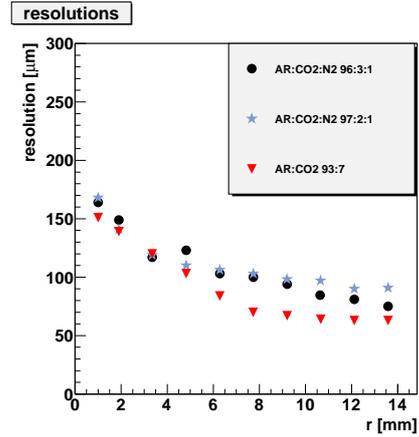}}
	 \caption{Single tube resolution of the standard gas (triangles) and two gas mixtures (stars and circles) containing 1 \% N$_{2}$}
	 \end{figure}

\section{MDT performance under fast n background}

	 To study the dependence of the single tube resolution on fast neutron background, we used the test setup shown in
	 Fig.\,3,
	 at the 14 MV tandem of the Munich Maier-Leibnitz Laboratory (MLL).
	 Almost monoenergetic 11 MeV neutrons were produced bombarding a hydrogen gas target with 60 MeV $^{11}$B ions.
	 The small test chamber (3 layers, 3 tubes per layer), surrounded by SI-strip detectors for reference tracks and scintillation 
	 counters for cosmic trigger, was operated under standard gas and electronics conditions. The Si-strip detectors have a
	 position resolution of better than 20 $\mu m$.\\
	 \begin{figure}[h]
	 \includegraphics[width=5.0cm]{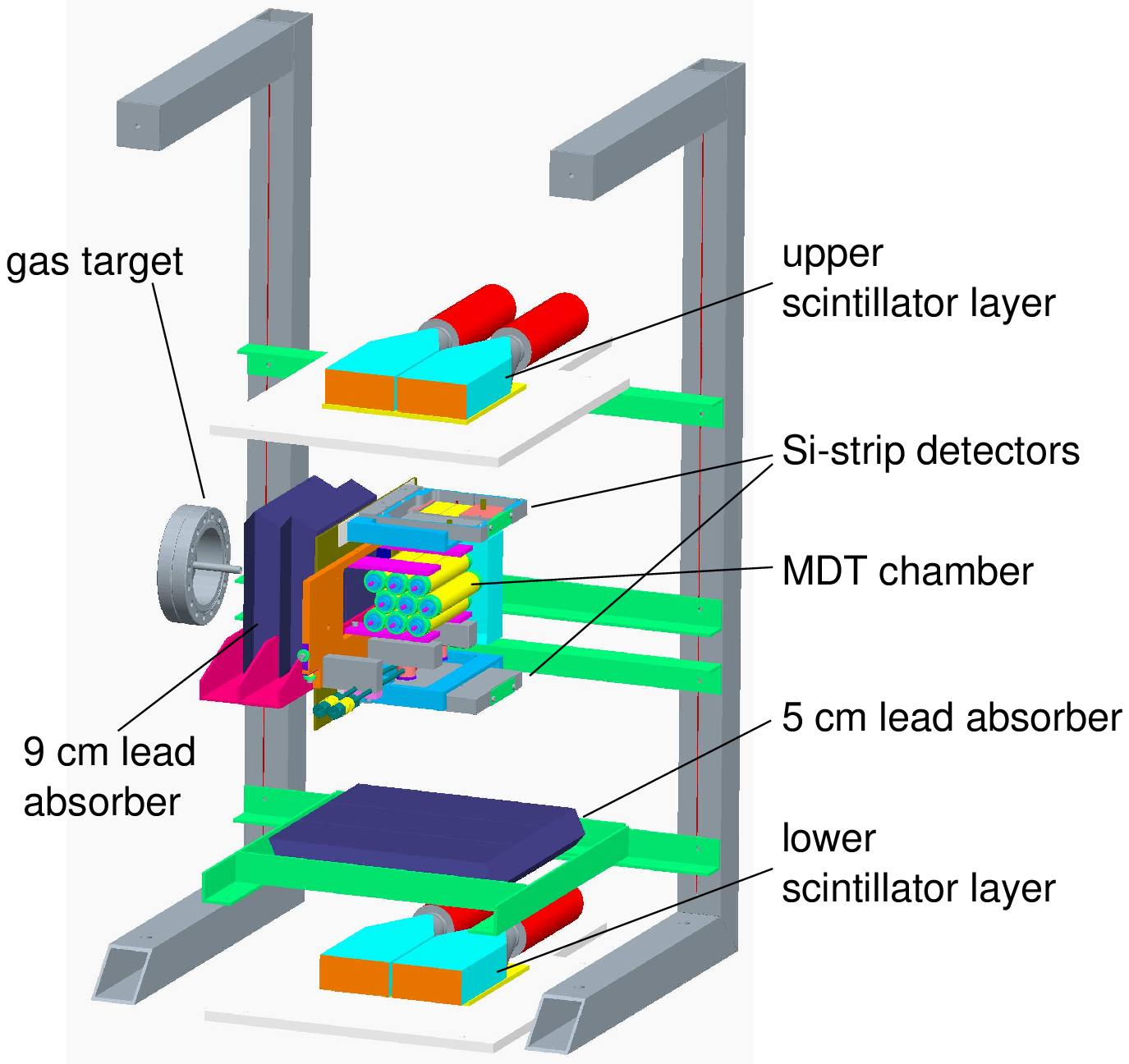}
	 \includegraphics[width=3.7cm]{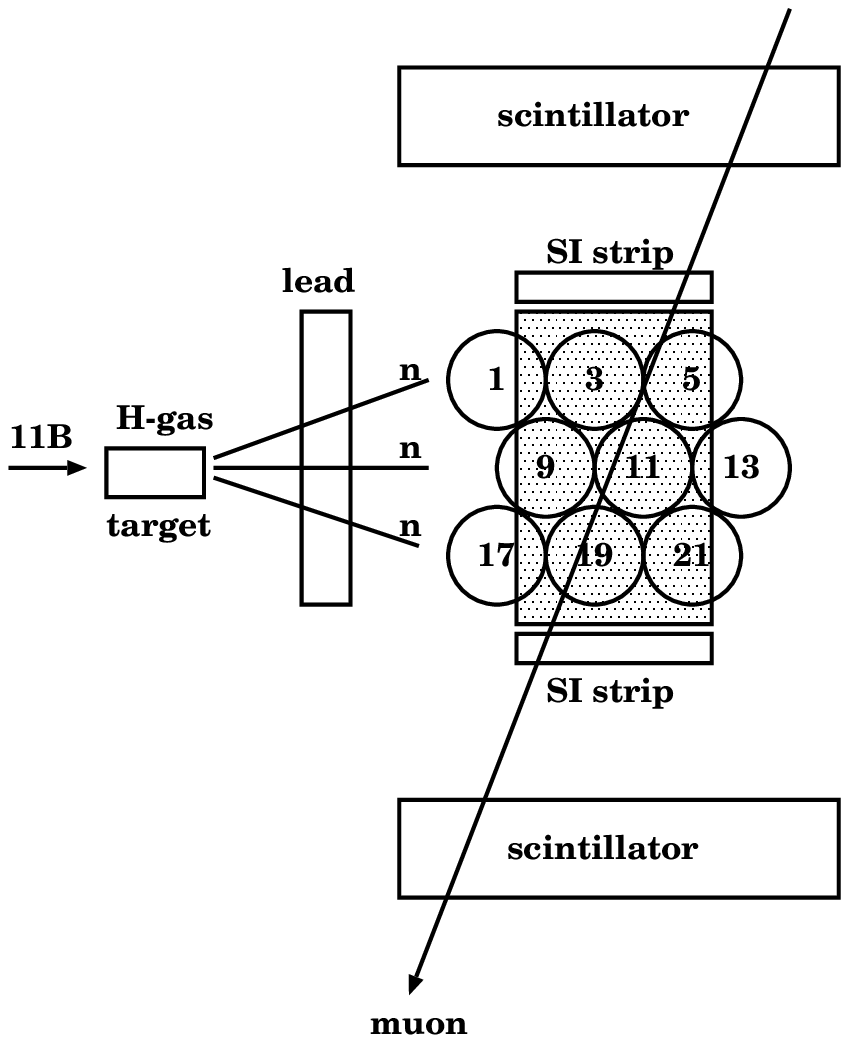}
	 \caption{Setup to study a small MDT chamber under neutron irradiation.}
	 \end{figure}
	 The single tube resolution was determined in runs with and without background neutrons. The result for the first method, comparing the measured
	 radii for tubes 3, 9, 11, 19, 21 (see Fig.\,4) with the tracks through the SI-strip detectors, is shown in the lower part of Fig.\,4. 
	 The upper part of Fig.\,4 shows
	 the result for the comparison of a track fitted to a triplet of tubes with the individual drift radii per tube.
	 
	 The degradation of the single tube resolution under 11 MeV neutron flux densities of 4\,-\,16 $\frac{kHz}{cm^{2}}$
	 is about 10 $\mu m$. Using triple sums, in most cases a small radius contributes to the calculation, worsening the resolution as observed in Fig.\,2.\\ 
	 \begin{figure}[h]
	 \centerline{\includegraphics[width=5.5cm]{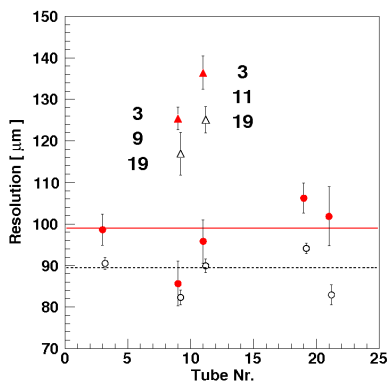}}
	 \caption{Single tube resolution with (filled markers) and without (open markers) 11 MeV neutron background. The triangles show the results for the triple
	 sum method and the circles for the method using SI-reference detectors.}
	 \end{figure}
	 To determine the overall neutron efficiency, which quantifies the sensitivity of the MDT chamber to neutrons, the MDT chamber readout
	 was randomly triggered. Dividing the number of counts per second in a tube by the neutron flux leads to an efficiency
	 of $(4.0^{+1.6}_{-0.3})\cdot 10^{-4}$. This value is indicated in Fig.\,5, where two simulations\cite{Baranov} are shown as well.
	 \begin{figure}[h]
	 \centerline{\includegraphics[width=5.5cm]{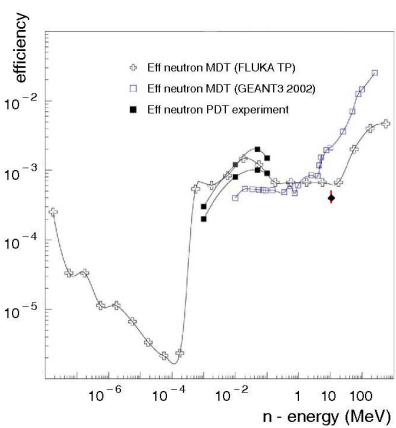}}
	 \caption{Different simulations of neutron efficiency\cite{Baranov}. Our result is the data point at 11 MeV and $4\cdot10^{-4}$.}
	 \end{figure}

\section{Simulations of drift tubes under $\gamma$\,-\,background}

	 $\gamma$\,-\,background events as well are ionizing the counting gas. The produced primary electrons drift towards the anode wire and close
	 to the wire due to the gas avalanche most of the electron-ion pairs are produced. The ions (the ring in Fig.\,6, left plot) drift in a time scale 
	 of 4 ms to the tube wall.
	 \begin{figure}[h]
	 \includegraphics[width=4cm]{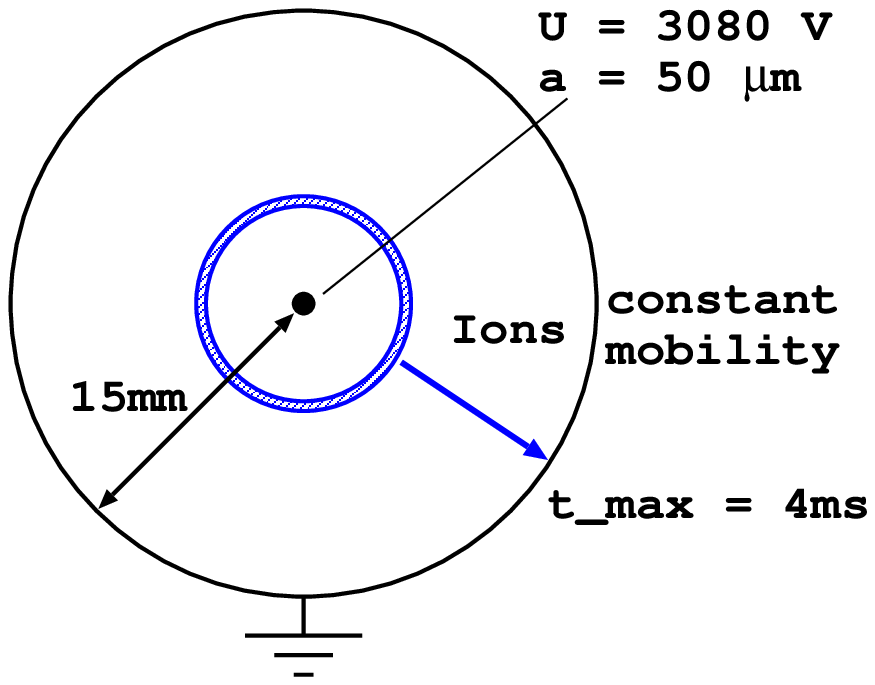}
	 \includegraphics[width=4cm]{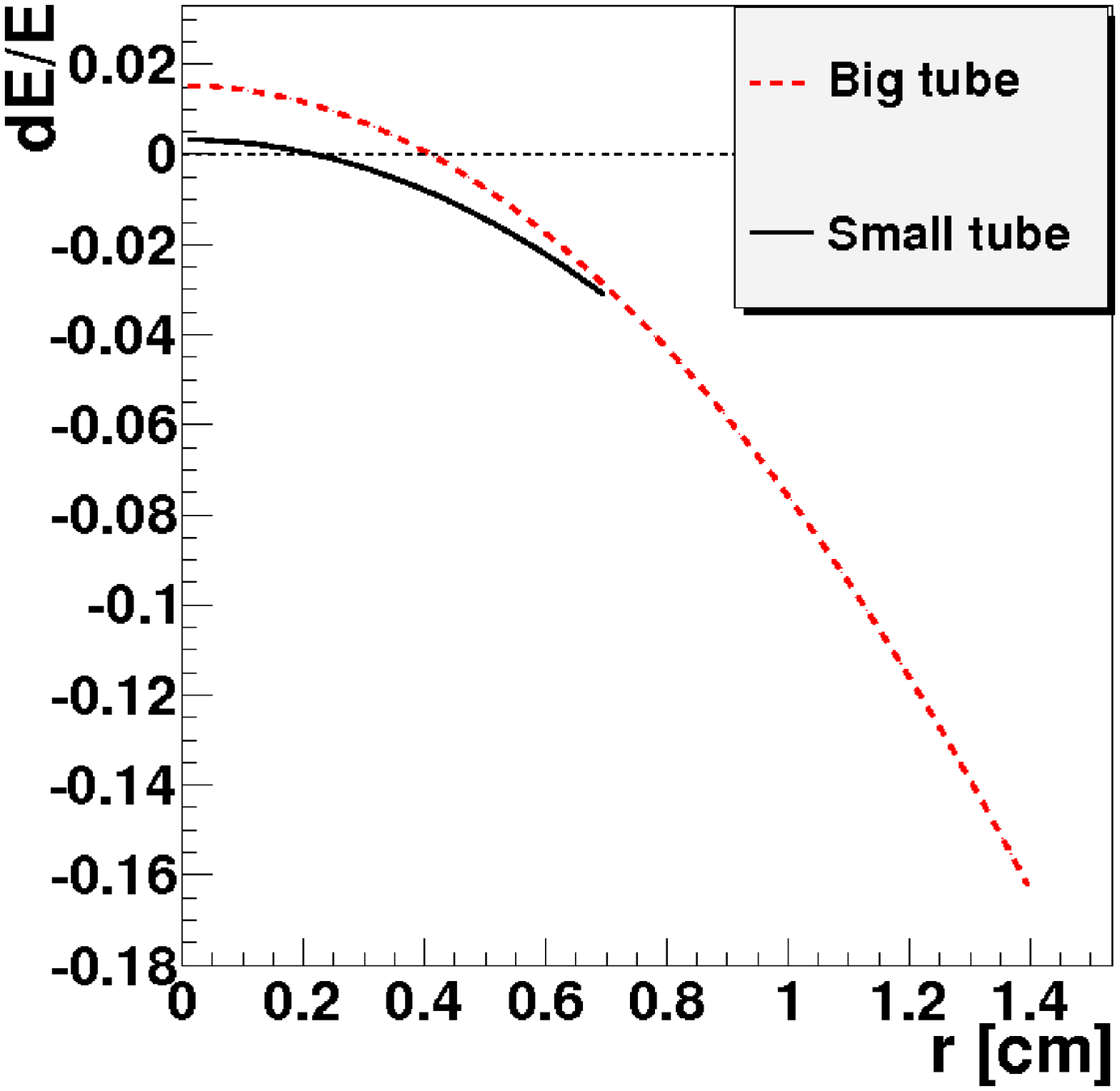}
	 \caption{Left: Drift tube with space charge; Right: Relative change of the electric field due to space charge effects.}
	 \end{figure}
	 These space charges contribute to the total drift field. For big radii the field is enhanced and one gets faster drift times, whereas
	 the field reduction at small radii lead to less gas gain. Both effects account for the degradation of the tube resolution.
	 
	 Fig.\,6 (right plot) shows the relative change of the electric field due to space charges. Using tubes with a diameter of 15 mm will
	 reduce the sensitivity on $\gamma$\,-\,background considerably.
	 
	 Garfield simulations\cite{Garfield} of the tube resolution including the field change and gain drop confirm measured results of the dependence of
	 the resolution on the $\gamma$\,-\,background count rate for 30\,mm drift tubes (see Fig.\,7). 
	 \begin{figure}[h]
	 \centerline{\includegraphics[width=5.5cm]{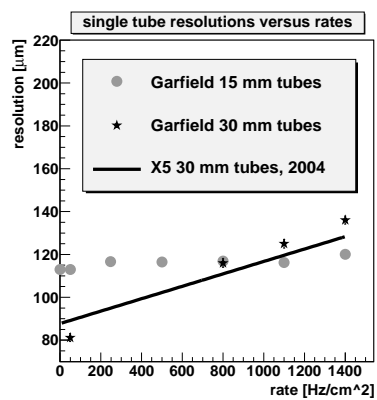}}
	 \caption{Comparison of simulated (circles and stars) and a fit (line) to measured resolutions of 30 mm tubes.}
	 \end{figure}
	 The simulation predicts that at counting rates of 1400 $\frac{Hz}{cm^{2}}$ the resolution for small tubes is only minorily deteriorated.
	 
%\appendix

%
%
%===============================================================
%
%   References
%
%===============================================================
%
%


\begin{thebibliography}{00}

%\bibitem{}
%
	 \bibitem{ATLASTechreport}
	 \newblock{ATLAS Collaboration: Technical Design Report for the ATLAS Muon Spectrometer, CERN/LHCC97-22, May 1997.}

	 \bibitem{Aleksadiss}
	 \newblock{M. Aleksa et al.: MDT Performance in a High Rate Background Environment , ATL-MUON-98-258, 1998.}
  
	 \bibitem{Deile1}
	 \newblock{M. Deile et al.: Performance of the ATLAS Precision Muon Chambers under LHC Operating
	 Conditions, Nucl. Instr. and Meth. \textbf{A518} (2004) 65.}

	 \bibitem{Deile2}
	 \newblock{M. Deile et al.: Resolution and Efficiency of the ATLAS Muon Drift-Tube Chambers at
	 High Background Rates, Nucl. Instr. and Meth. \textbf{A535} (2004) 212.}

	 \bibitem{Horvat}
	 \newblock{S. Horvat et al.: Operation of the ATLAS Muon Drift-Tube Chambers at High Background
	 Rates and in Magnetic Fields, IEEE Transactions on Nuclear Science, vol. 53, no. 2, pp.
	 562-566, 2006.}

	 \bibitem{Garfield}
	 \newblock{http://garfield.web.cern.ch/garfield/}

	 \bibitem{Baranov}
	 \newblock{S. Baranov et al.: Estimation of Radiation Background, Impact on Detectors, Activation and Shielding Optimization in ATLAS, ATL-GEN-2005-001, CERN 2005.}

	 \bibitem{Aleksa}
	 \newblock{M. Aleksa: Performance of the ATLAS Muon Spectrometer, Dissertation, Vienna 1999.}

\end{thebibliography}
\end{document}